\documentclass[10pt,showpacs,showkeys,a4paper,groupedaddress]{revtex4}
\usepackage{graphicx}
\usepackage{tabularx}
\usepackage{amsmath}
\usepackage{amsfonts}
\usepackage{amssymb}
\usepackage{natbib}

\begin{document}
\title{Surface-wave interferometry on single subwavelength slit-groove structures fabricated on gold films}
\date{\today}
\author{F. Kalkum}\altaffiliation{Permanent address: Physikalisches
Institut, Universit{\"a}t Bonn, Wegelerstrasse 8, 53115 Bonn, Germany}\affiliation{IRSAMC/LCAR\\
Universit\'e Paul Sabatier, 118 route de Narbonne,\\31062
Toulouse, France}\author{G. Gay, O. Alloschery, J. Weiner}
\affiliation{IRSAMC/LCAR\\
Universit\'e Paul Sabatier, 118 route de Narbonne,\\31062
Toulouse, France}\email{jweiner@irsamc.ups-tlse.fr}
\author{H. J. Lezec}\affiliation{Centre National de la Recherche Scientifique, 3,
rue Michel-Ange, 75794 Paris cedex 16, France}\affiliation{Thomas
J. Watson Laboratories of Applied Physics, California Institute of
Technology, Pasadena, California 91125 USA} \author{Y. Xie, M.
Mansuripur}\affiliation{College of Optical Sciences, University of
Arizona, Tucson, Arizona 85721 USA}


\begin{abstract}
We apply the technique of far-field interferometry to measure the
properties of surface waves generated by two-dimensional (2D)
single subwavelength slit-groove structures on gold films. The
effective surface index of refraction $n_{\mathrm{surf}}$ measured
for the surface wave propagating over a distance of more than
12~$\mu$m is determined to be $n_{\mathrm{surf}}=1.016\pm 0.004$,
to within experimental uncertainty close to the expected bound
surface plasmon-polariton (SPP) value for a Au/Air interface of
$n_{\mathrm{spp}}=1.018$.  We compare these measurements to
finite-difference-time-domain (FDTD) numerical simulations of the
optical field transmission through these devices.  We find
excellent agreement between the measurements and the simulations
for $n_{\mathrm{surf}}$.  The measurements also show that the
surface wave propagation parameter $k_{\mathrm{surf}}$ exhibits
transient behavior close to the slit, evolving smoothly from
greater values asymptotically toward $k_{\mathrm{spp}}$ over the
first 2-3~$\mu$m of slit-groove distance $x_{\mathrm{sg}}$. This
behavior is confirmed by the FDTD simulations.
\end{abstract}



\maketitle

\section{Introduction}
Recent measurements \cite{GAV06b,GAV06a} have characterized
surface waves arising from optical excitation of a series of
subwavelength slit-groove structures fabricated on silver films.
The amplitude, wavelength and phase of these surface waves have
been measured over a slit-groove distance of a few microns. After
an initial rapid amplitude decrease extending to $\sim 3$~$\mu$m
 from the slit edge, the interference fringe persists over
several microns with a near-constant amplitude and a contrast
$\simeq 0.3$. The ``near-zone" of rapidly changing surface wave
character indicates initial transient phenomena, while the
longer-range ``far-zone" settling to constant amplitude and
contrast is the signature of a guided mode surface plasmon
polariton (SPP). The measured interference fringe wavelength
$\lambda_{\mathrm{surf}}=814\pm 8$~nm resulted in the
determination of an effective index of refraction
$n_{\mathrm{surf}}=1.047$, significantly higher than expected from
conventional theory\,\cite{Raether}, $n_{\mathrm{spp}}=1.015$. The
question naturally arose as to whether this result is simply a
consequence of interference fringe sampling over an interval
predominantly in a transient near-zone peculiar to silver films,
or was related to the specific surface properties of the silver
films deposited on fused silica\,\cite{LH06}, or may indicate a
more \emph{generic} phenomenon related to the transient properties
of surface waves, generated by subwavelength slits, within 2-3
wavelengths of the edge of origin. Subsequent investigation of the
physical-chemical properties of the silver films confirmed that
they were free of surface contaminants, and FDTD simulations
showed excellent agreement with the experimental results in silver
films\,\cite{GAW06,GAW06a}.

In order to explore these surface waves further we have carried
out a series of experiments similar to those already reported but
using evaporated gold films instead of silver. We performed only
``output-side" experiments\,\cite{GAV06b} because they yield a
phase modulated interference fringe less susceptible to noise than
the amplitude modulation signal of the ``input-side"
experiments\,\cite{GAV06a}. In addition we have compared the
measurements to detailed field amplitude and phase maps generated
by FDTD numerical solutions \cite{XZM04,XZM05} to the Maxwell
equations in the vicinity of the slit-groove structure. Both
experiments and FDTD simulation show that the surface wave
exhibits transient properties in wavelength and amplitude in the
near-zone. This transient behavior can be interpreted in terms of
surface modes all of which dissipate beyond the near-zone except
for $k_{\mathrm{spp}}$, the bound surface mode.

\section{Experimental Setup}
The structures were fabricated with a focused ion beam (FIB)
station as described previously in \cite{GAV06a}. The experiments
were performed on the same home-built goniometer setup used in the
silver studies.  For convenience, Figs.\,1,2 of \cite{GAV06b} are
reproduced here to show the principle of the measurement and the
schematic arrangement of the apparatus.
\begin{figure}
\centering
\includegraphics[width=0.45\columnwidth]{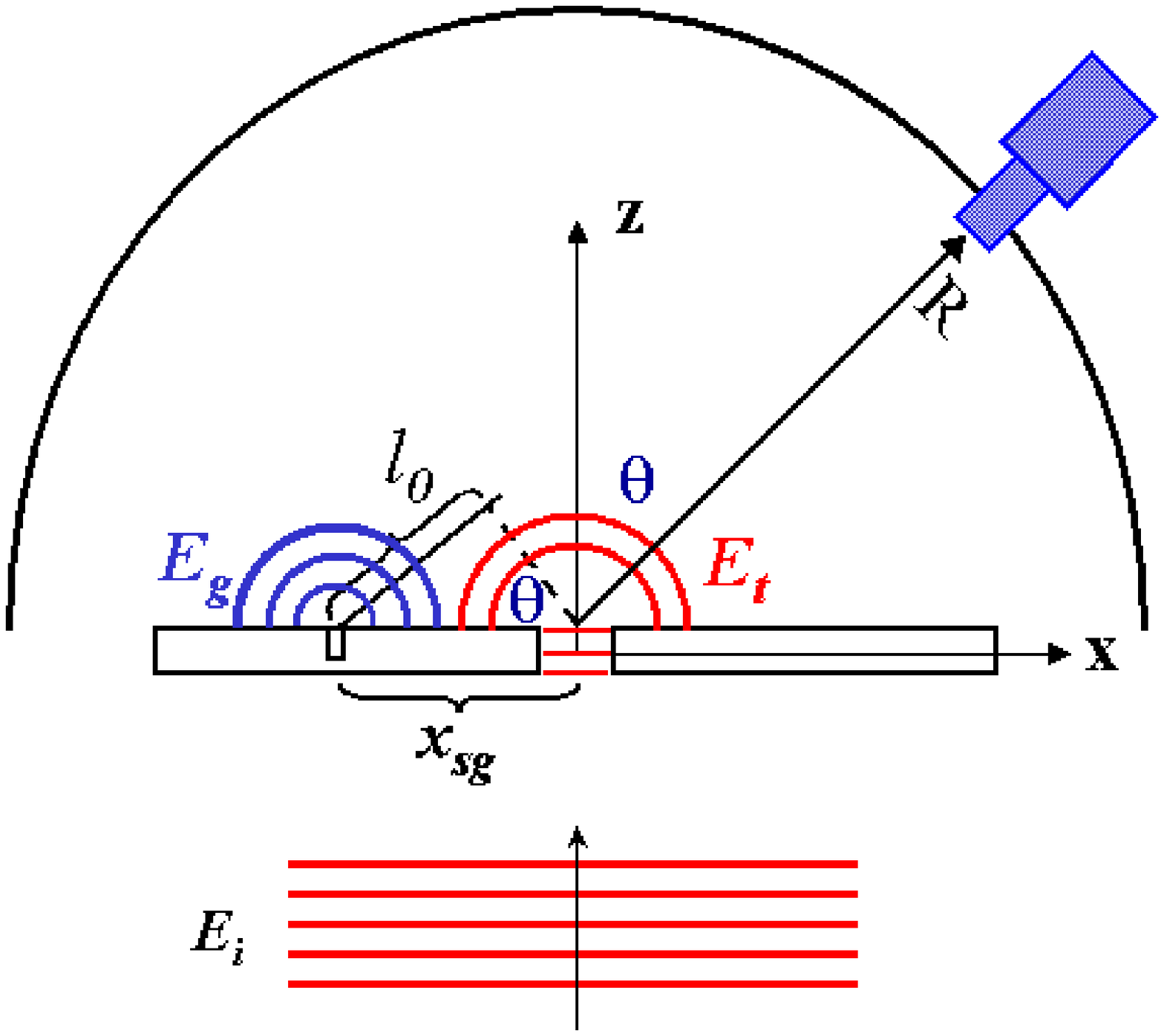}\caption{Diagram
showing interfering wavefronts, optical path difference between
$E_t$ and $E_g$, and far-field detection. Slit dimensions are
100~nm width, 20~$\mu$m length. Groove dimensions are 100~nm
width, 100~nm nominal depth, 20~$\mu$m length.  The evaporated
gold layer deposited on a 1~mm fused silica substrate has a 400~nm
nominal thickness.}\label{Fig:InterferenceDiag}%
\bigskip
\centering
\includegraphics[width=0.45\columnwidth]{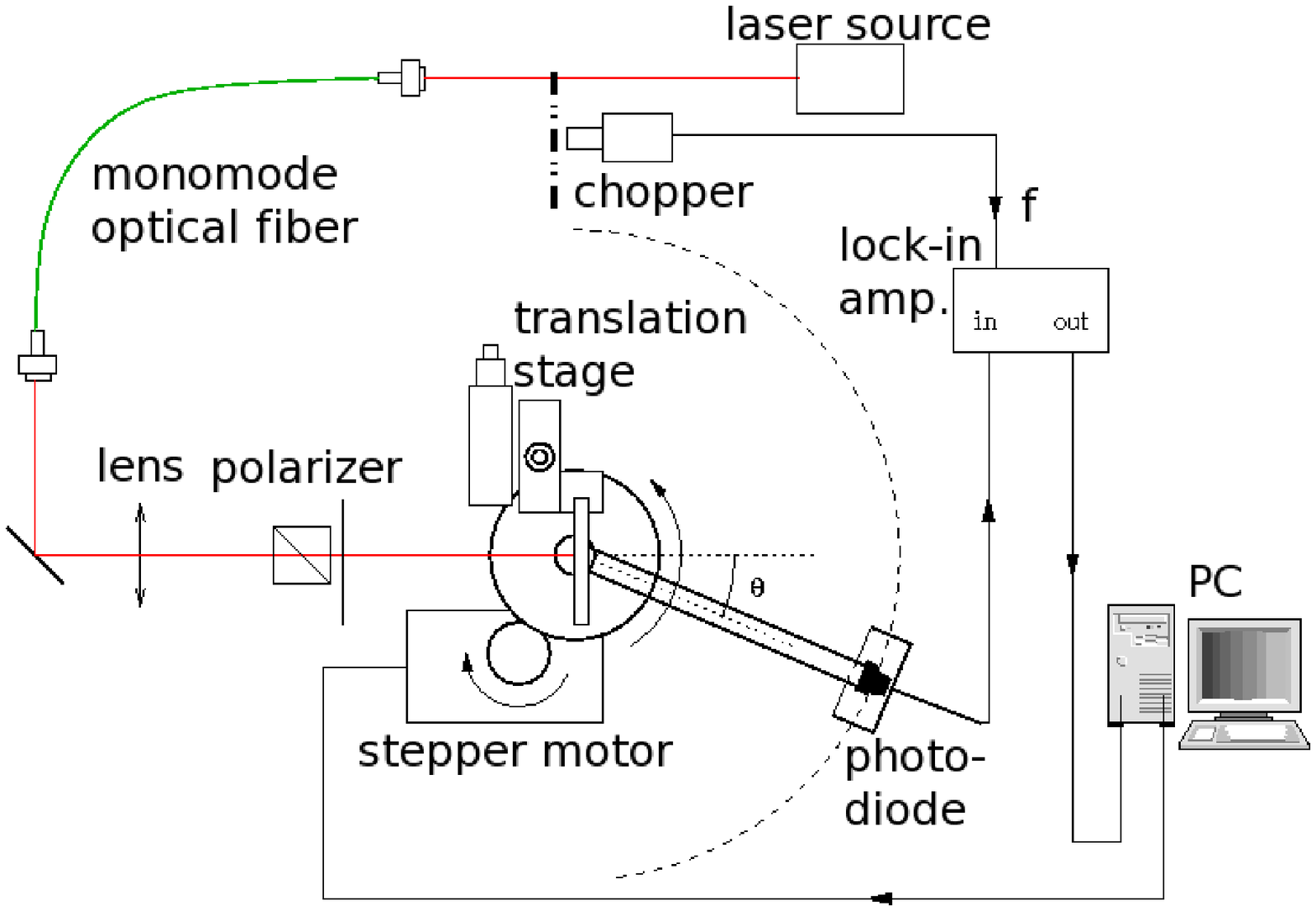}\caption{Goniometer setup for measuring
far-field light intensity and angular distributions.
 See text for description.}\label{Fig:goniometre}
\end{figure}
Output from a temperature stabilized diode laser source, is
modulated at 850 Hz by a mechanical chopper, injected into a
monomode optical fibre, focused and linearly polarized (TM
polarization, H-field parallel to the slit long axis) before
impinging perpendicularly on the matrix of structures mounted in a
sample holder.  The beam waist diameter and confocal parameter of
the illuminating source are 300~$\mu$m and 33~cm, respectively.
The sample holder itself is fixed to a precision x-y translator,
and individual slit-groove structures of the 2-D matrix are
successively positioned at the laser beam waist.  A photodiode
detector is mounted at the end of a 200 mm rigid arm that rotates
about the center of the sample holder.  A stepper motor drives the
arm at calibrated angular increments of 1.95 mrad per step, and
the overall angular resolution of the goniometer is $\simeq 4$
mrad. The photodetector output current passes to a lock-in
amplifier referenced to the optical chopper wheel. Data are
collected on a personal computer that also controls the goniometer
drive.
\section{Results and Analysis}\subsection{Measurements}
With the detector rotated perpendicular to the structure plane
($\theta = 0$ in
Figs.\,\ref{Fig:InterferenceDiag},\,\ref{Fig:goniometre}) the
expression for the normalized detected intensity $I/I_0$ as a
function of slit-groove distance $x_{\mathrm{sg}}$ is given by
\begin{equation}
\frac{I}{I_0}\propto 1
+\eta_o^2+2\eta_o\cos(k_xx_{\mathrm{sg}}+\varphi_{\mathrm{int}})\label{Eq:fringes}
\end{equation}
where $\eta_0$ is related to the fringe contrast $C$ through
\begin{equation}
C=\frac{2\eta_o}{1+\eta_o^2}\label{Eq:contrast}
\end{equation}
In the argument of the cosine term
$k_x=2\pi/\lambda_{\mathrm{eff}}$ relates the propagation
parameter of the surface wave to the effective surface wavelength,
and the ``intrinsic phase" $\varphi_{\mathrm{int}}$ is any phase
contribution not directly due to the propagation path length
$x_{\mathrm{sg}}$.  It may be associated with phase shifts at the
slit or groove structures. The interferometry measurements were
carried out on 4 separate substrates, each substrate containing
about 50 structures in which the the slit-groove separation was
systematically varied from 50~nm to more than 12~$\mu$m in
increments of 50~nm.
\begin{figure}
\centering
\includegraphics[width=0.49\columnwidth]{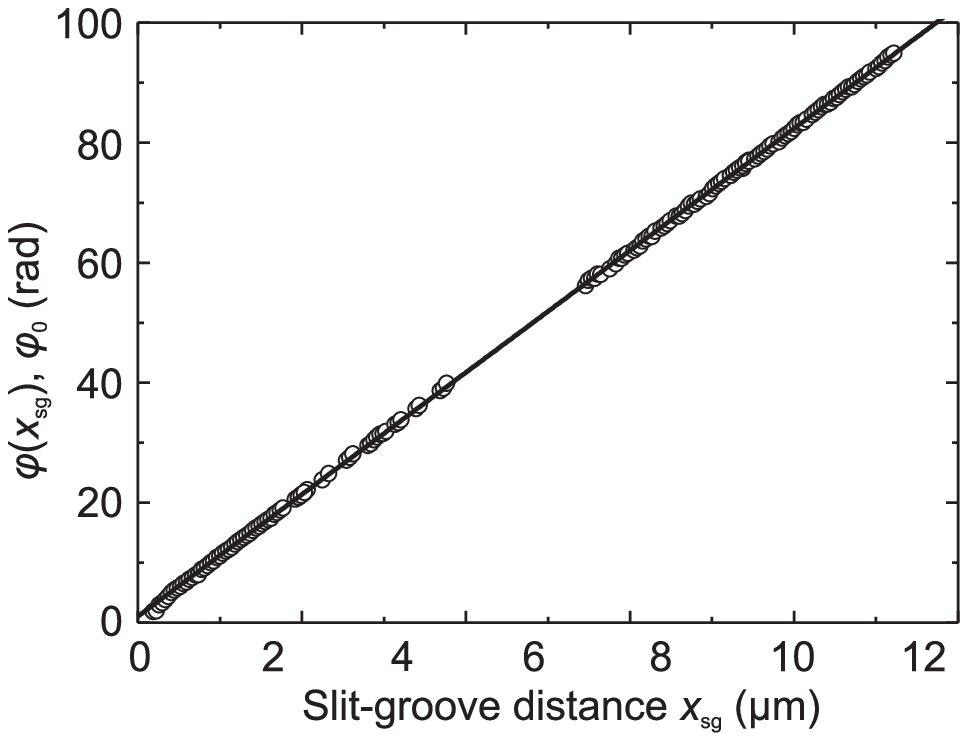}
\centering
\includegraphics[width=0.49\columnwidth]{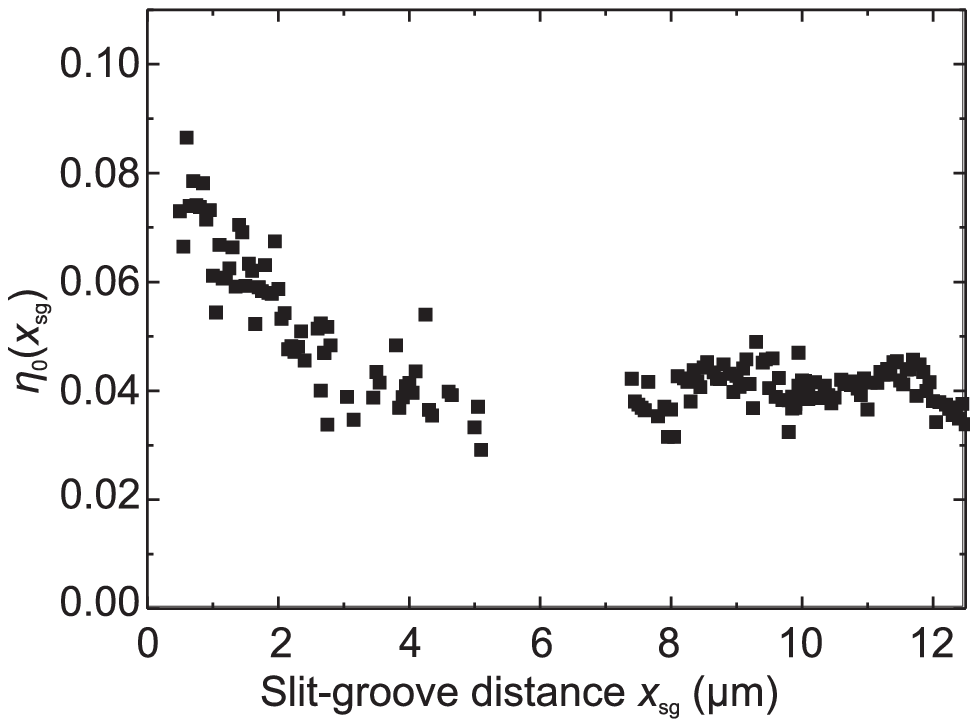}\caption{Left panel: Points are the measured fringe
phase $\varphi(x_{\mathrm{sg}})$ as a function of slit-groove
distance $x_{\mathrm{sg}}$. The straight-line fit is
$\varphi_0=k_{\mathrm{surf}}x_{\mathrm{sg}}+\varphi_{\mathrm{int}}$
with constant slope $k_{\mathrm{surf}}$ and intercept
$\varphi_{\mathrm{int}}$.  Gap in the data in left and right
panels of this figure and in the left and right panels of
Fig.\,\ref{Fig: phase-error-detail} is due to defective structures
in this interval. Right panel: Fringe amplitude $\eta_0\simeq 1/2
C$ as a function of $x_{\mathrm{sg}}$ where $C$ is the
interference fringe contrast.}\label{Fig: contrast plot}
\end{figure}
The left panel of Fig.\,\ref{Fig: contrast plot} plots the
measured interference fringe phase against the slit-groove
distance $x_{\mathrm{sg}}$. The fitted value for
$k_{\mathrm{surf}}=2\pi/\lambda_{\mathrm{surf}}$ determines the
effective surface index of refraction $n_{\mathrm{surf}}$, and
extrapolation to zero slit-groove distance $x_{\mathrm{sg}}$
determines the intrinsic phase $\varphi_{\mathrm{int}}$. The fit
from the left panel of Fig.\,\ref{Fig: contrast plot} yields
\begin{equation}
n_{\mathrm{surf}}=\frac{\lambda_0}{\lambda_{\mathrm{surf}}}=1.016\pm
0.004\qquad\mbox{and}\qquad\varphi_{\mathrm{int}}=0.35\pi \pm
0.01\pi\label{Eq: n-effective}
\end{equation}
The right panel of Fig.\,\ref{Fig: contrast plot} plots $\eta_0$,
the amplitude factor of the interference term in
Eq.\,\ref{Eq:fringes}, as a function of slit-groove distance. From
Eq.\,\ref{Eq:contrast} this amplitude factor can be expressed in
terms of the fringe visibility or contrast $C$ as
\begin{equation}
\eta_0=\frac{1-\sqrt{1-C^2}}{C}\simeq\frac{1}{2}C,\qquad C\ll1
\label{Eq:eta-contrast}
\end{equation}
\begin{figure}
\centering
\includegraphics[width=0.49\columnwidth]{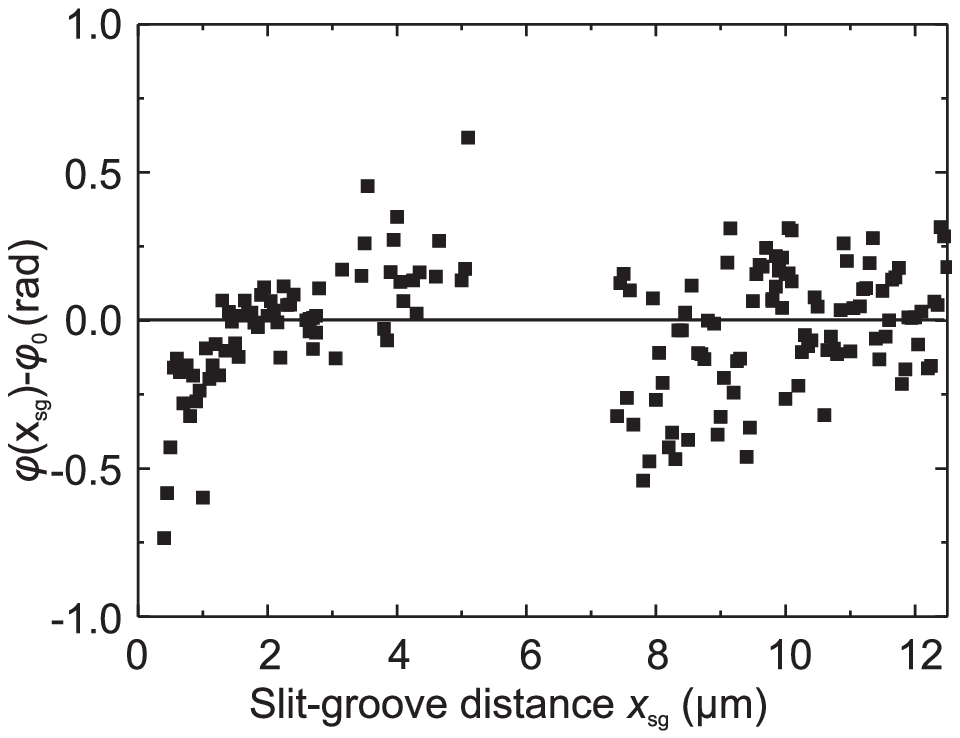}
\hfill
\centering
\includegraphics[width=0.49\columnwidth]{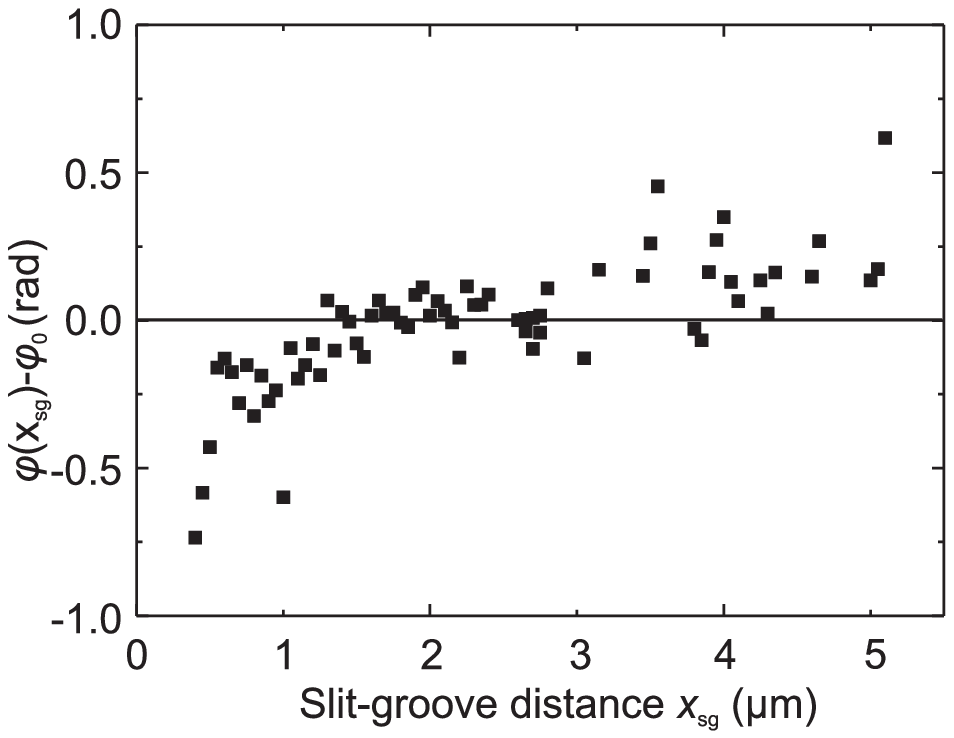}
\caption{Left panel: Fringe phase difference
$\varphi(x_{\mathrm{sg}})-\varphi_0$ as a function of slit groove
distance $x_{\mathrm{sg}}$. Deviation in the near-zone from
$\varphi_0$ indicates that early, transient fringe oscillation is
slightly greater and approaches $\varphi_0$ asymptotically in the
far-zone beyond $\sim 2~\mu$m slit-groove distance.  Right panel:
Same data as shown in left panel
but on an expanded scale of slit-groove distances to emphasize the
curvature in $\varphi(x_{\mathrm{sg}})-\varphi_0$ in the
near-zone.}\label{Fig: phase-error-detail}
\end{figure}
Although the fringe contrast shown in the right panel of
Fig.\,\ref{Fig: contrast plot} is about a factor of 5 below that
measured for silver structures\,\cite{GAV06b}, the same rapid
fall-off in the near-zone, $x_{\mathrm{sg}}\simeq 0-3~\mu$m,
followed by a near-constant contrast beyond is observed.  This
contrast behavior is evidence of surface-wave transient phenomena
in the near-zone. More evidence of this transient behavior is
shown in left and right panels of Fig.\,\ref{Fig:
phase-error-detail} that plot $\varphi(x)-\varphi_0$ vs.
$x_{\mathrm{sg}}$, where
$\varphi_0=k_{\mathrm{surf}}x_{\mathrm{sg}}+\varphi_{\mathrm{int}}$
is the best-fit linear trace in the left panel of Fig.\,\ref{Fig:
contrast plot} over the range of slit-groove distances out to
12~$\mu$m. A pronounced departure from the asymptotic value of
$\varphi_0$ is evident in the near-zone of slit groove distances,
indicating that the fringe oscillation frequency is initially
somewhat greater than the SPP value and smoothly decreases to it
beyond the near-zone.

\subsection{Numerical simulations}
The time-dependent Maxwell equations are solved numerically using
an FDTD non-conformal grid refinement method in Cartesian space
coordinates.  The methodology is described in greater detail in
Refs\,\cite{ZMM04,XZM04,XZM05}.

\begin{figure}
\centering
\includegraphics[width=0.75\columnwidth]{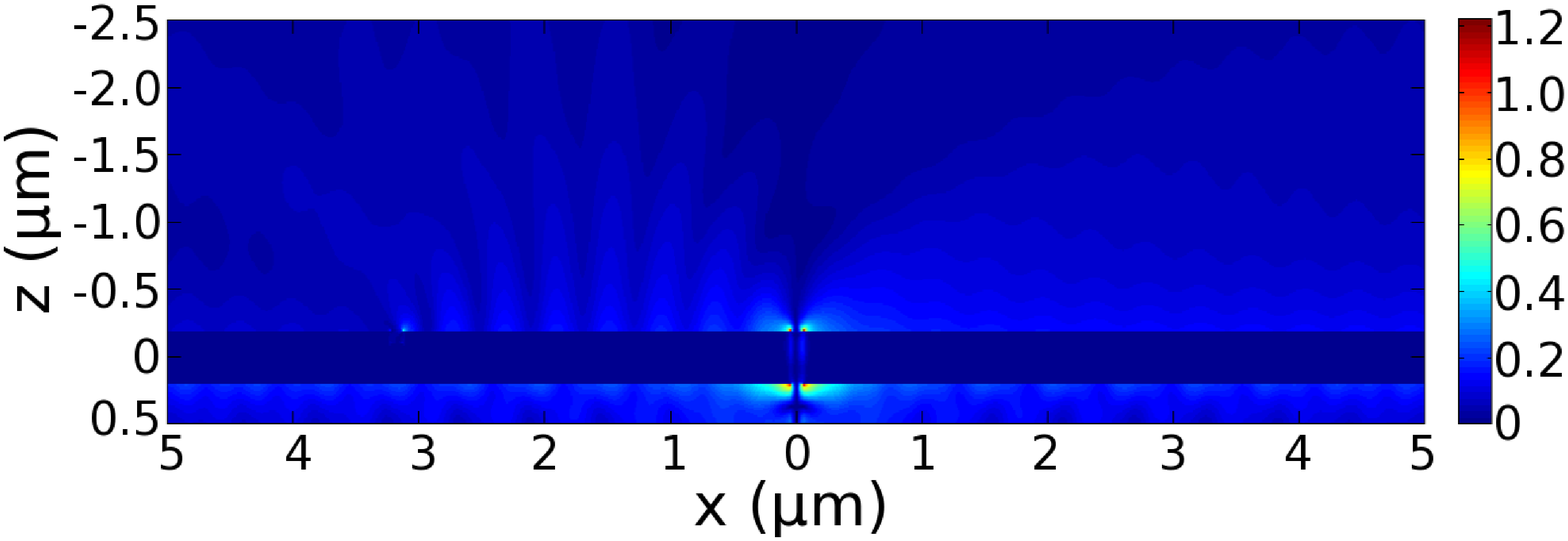}\caption{FDTD simulations for slit-groove
center-to-center distance of 3.18~$\mu$m, slit and groove widths
100~nm, groove depth 100~nm and gold film thickness 400~nm. Map
shows $|\mathrm{E_z}|$, z-components (perpendicular to input and
output facets) of the electric field amplitude in the vicinity of
the input and output surfaces.}\label{Fig: Ez}
\medskip
\centering
\includegraphics[width=0.75\columnwidth]{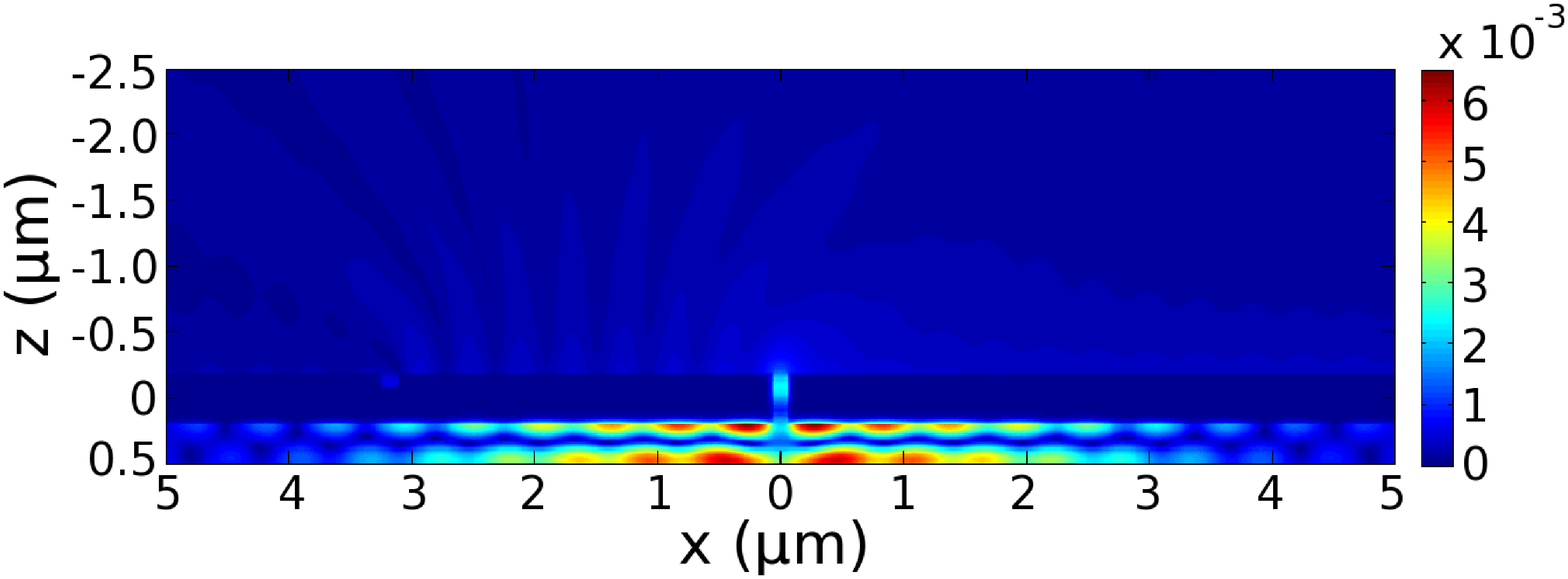}\caption{FDTD simulations for slit-groove
center-to-center distance of 3.18~$\mu$m, slit and groove widths
100~nm, groove depth 100~nm and gold film thickness 400~nm. Map
shows $|\mathrm{H_y}|$, y-components (parallel to the slit and
groove long axis) of the magnetic field amplitude in the vicinity
of the input and output surfaces.}\label{Fig: Hx}
\end{figure}

Figure \ref{Fig: Ez} shows a field map of the z-component of the
electric field amplitude, and Fig.\,\ref{Fig: Hx} shows a field
map of the y-component of the magnetic field amplitude.  The
$|\mathrm{E_z}|$ map clearly shows the dipolar charge distribution
concentrated at the corners of the slit on the input and output
planes of the structure. These corner charge concentrations result
from currents induced on the input side of the gold film by the
magnetic field components $|\mathrm{H_y}|$ shown in
Fig.\,\ref{Fig: Hx}. The incident light propagates from below
through the fused silica substrate onto the gold film and through
the 100~nm wide, 400~nm thick slit. The incident light is TM
polarized and the guided mode propagating along the $\pm z$
direction within the slit sets up a standing wave resulting in a
high $|\mathrm{E_z}|$, $|\mathrm{H_y}|$ amplitudes at the output
plane. The groove is at the output side of the gold film; and, in
the simulations depicted, the distance between the center of the
slit and the center of the groove is 3.18~$\mu$m. The absolute
value of the z-component of the electric field amplitude
$|\mathrm{E_z}|$ just above (or just below) the gold film is
proportional to the surface charge density at the film surface.
Note that on the output side surface, in the region between the
groove and the slit, the surface wave excited at the left edge of
the slit travels to the groove, is reflected from the groove's
right edge, then interferes with itself. The standing wave is
clearly visible on and near the output side surface. Within the
slit, on the vertical walls, $\mathrm{E_z}$ is fairly strong as
well. Here, however, $\mathrm{E_z}$ is parallel to the metallic
surface, and its presence within the skin depth of the slit walls
does not signify the existence of surface charges; instead, the
$\mathrm{E_z}$-field in this region is responsible for the surface
currents that carry the charges back and forth between the
entrance and exit facets of the slit.  Figure \ref{Fig: Hx} shows
the magnitude of the magnetic field $|\mathrm{H_y}|$ over the same
region as Fig.\,\ref{Fig: Ez}. Interference fringes between the
incident and reflected beams on fused silica substrate are clearly
visible. Note also the interference fringes between the excited
evanescent waves (mainly SPP) and the incident beam on the
entrance facet of the gold film adjacent the substrate. Inside the
slit, $|\mathrm{H_y}|$ shows a dark band; this is caused by
interference between the upward-moving guided mode within the slit
and the reflected, downward-travelling mode.

In addition to the electric and magnetic field components
$\mathrm{E_z,H_y}$ at the surface, the light transmission
efficiency in the $z$ direction through the slit as a function of
slit-groove distance $x_{\mathrm{sg}}$ was calculated and is shown
in Fig.\,\ref{Fig: TvsD}. The transmission efficiency $T$ is
defined as the ratio of the $z$-component of the Poynting vector
$S_z^{out}$ on the output side, integrated over $x$, to the total
energy flux incident on the slit $S_z^{in}$.
\medskip
\begin{figure}\centering
\includegraphics[width=0.50\columnwidth]{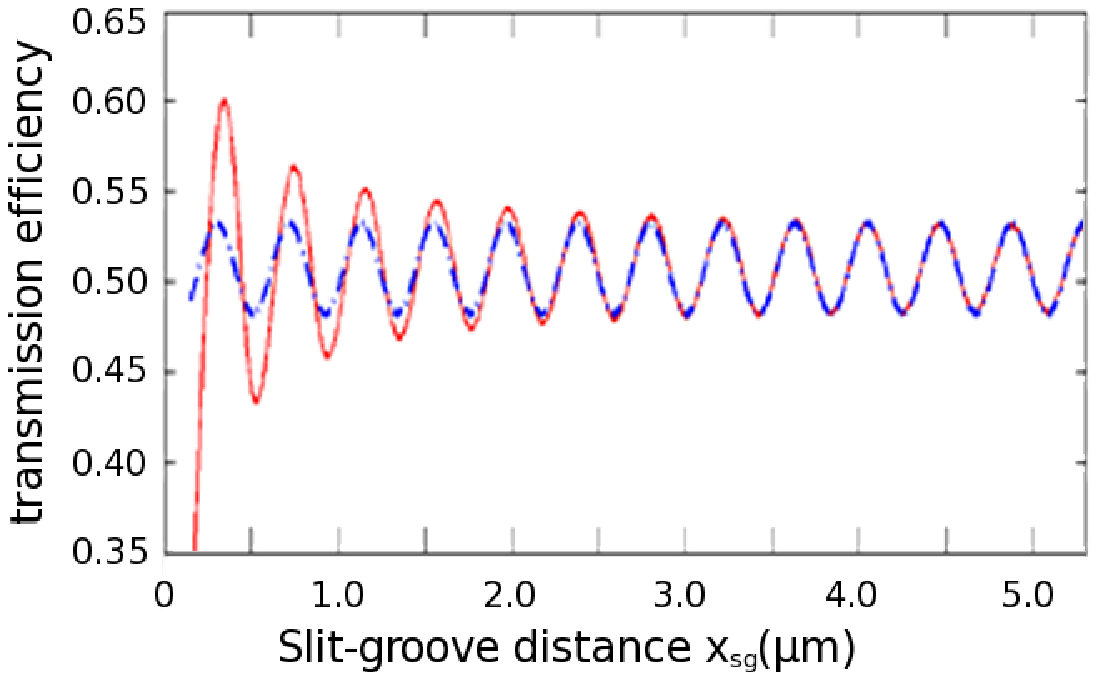}\caption{FDTD calculations of the transmission
efficiency $T=S_z^{out}/S_z^{in}$ as a function of
$x_{\mathrm{sg}}$. Red curve traces $T$, and the blue curve traces
a $\cos\left(2k_{\mathrm{surf}}^{\mathrm{fdtd}}\cdot
x_{\mathrm{sg}}+\varphi_{\mathrm{int}}^{\mathrm{fdtd}}\right)$ fit
to the oscillation in the asymptotic region.  Note the decreasing
transmission amplitude in the near-zone close to the slit edge and
the higher oscillation frequency compared to the asymptotic
harmonic wave.  Best-fit values for
$\lambda_{\mathrm{surf}}^{\mathrm{fdtd}}=2\pi/k_{\mathrm{surf}}^{\mathrm{fdtd}}=839$~nm
and intrinsic phase
$\varphi_{\mathrm{int}}^{\mathrm{fdtd}}=0.55\pi$~rad.}\label{Fig:
TvsD}
\bigskip
\includegraphics[width=0.50\columnwidth]{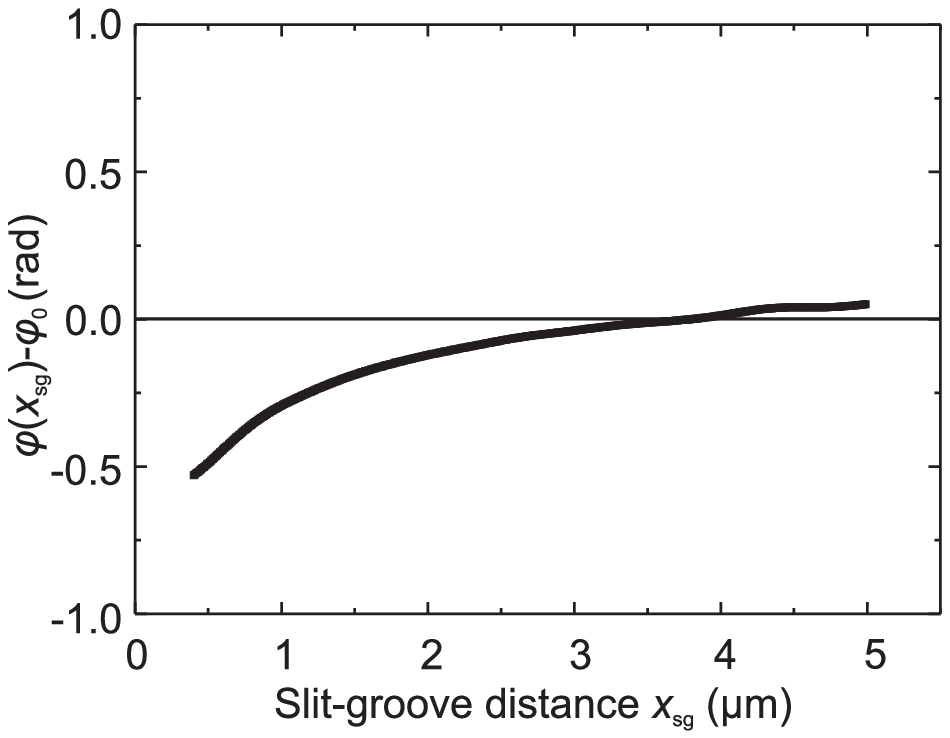}\caption{Phase difference $\varphi(x_{\mathrm{sg}})-\varphi_0$ as a
function of $x_{\mathrm{sg}}$, analogous to the right panel of
Fig.\,\ref{Fig: phase-error-detail} but derived from the FDTD
simulation data.
 Residual ``high frequency" oscillations in the phase difference, believed to be due to
 numeric artifacts in the FDTD results, have been smoothed.}\label{Fig:
calculated-phase-error}
\end{figure}
The red trace plots the transmission efficiency
$T=S_z^{out}/S_z^{in}$ as a function of $x_{\mathrm{sg}}$ at the
output plane.  As expected the transmission efficiency exhibits
pronounced oscillations with a rapid decrease in amplitude in the
near-zone followed by a constant amplitude oscillation out to
6~$\mu$m, the calculation limit.  In the far-zone the blue trace
fits the oscillations to a single cosine function,
$\cos\left[2\left(k_{\mathrm{surf}}^{\mathrm{fdtd}}\cdot
x_{\mathrm{sg}}+\varphi_{\mathrm{int}}^{\mathrm{fdtd}}\right)\right]$.
These oscillations, as can be seen in Figs.\,\ref{Fig: Ez}
and\,\ref{Fig: Hx}, arise from the superposition of waves launched
from the slit and back reflected at the groove.  Because these
waves are counterpropagating along $x$, rather than copropagating
along $z$, the intensity fringe frequency is twice the fringe
frequency of the far-field interferometry results expressed by
Eq.\,\ref{Eq:fringes}.  This standing wave at the output plane
results in
\begin{equation}
T=S_z^{out}/S_z^{in}\propto 2\left[1+\cos
2\left(k_{\mathrm{surf}}^{\mathrm{fdtd}}\cdot
x_{\mathrm{sg}}+\varphi_{\mathrm{int}}^{\mathrm{fdtd}}\right)\right]=
2\left\{1+\cos
2\left[\varphi\left(x_{\mathrm{sg}}\right)\right]\right\}
\label{Eq: SzStandingWave}
\end{equation}

Taking into account this factor of two in the argument of the
$\cos$ term, the fringe oscillation in the asymptotic far-zone is
in good agreement with measurement and the expected SPP. In the
near-zone the oscillation $\varphi\left(x_{\mathrm{sg}}\right)$
exhibits a definite ``chirp," and Fig.\,\ref{Fig:
calculated-phase-error} plots the deviation from the asymptotic
value $\varphi_0$ as function of $x_{\mathrm{sg}}$. Comparing
Fig.\,\ref{Fig: calculated-phase-error} to Fig.\,\ref{Fig:
phase-error-detail} we see that the FDTD results accord well with
deviations in the interferometric fringes measured in the far-
field.
\section{Discussion}
\subsection{Surface wave in the far-zone}
The index of refraction for the bound surface-plasmon-polariton
$n_{\mathrm{spp}}$ is given by the Raether formula\,\cite{Raether}
\begin{equation}
n_{\mathrm{spp}}=\sqrt{\frac{\epsilon_{\mathrm{m}}\epsilon_{\mathrm{d}}}{\epsilon_{\mathrm{m}}+\epsilon_{\mathrm{d}}}}\label{Fig:
Raether n}
\end{equation}
where $\epsilon_{\mathrm{m}}$ and $\epsilon_{\mathrm{d}}$ are the
real parts of the dielectric constants of metal and dielectric at
the interface on which the surface wave propagates.  Interpolation
of reflectivity data for gold\,\cite{JC72} at 852~nm yields
$\epsilon_{\mathrm{Au}}=-28.82$ and from Eq.\,\ref{Fig: Raether n}
the surface index of refraction for the surface plasmon polariton
at the gold-air interface is $\epsilon_{\mathrm{spp}}=1.018$. The
measured surface index of refraction reported here, to within
experimental uncertainty and in the far-zone, is in accord with
the SPP prediction.  The results from the FDTD calculations are
also in agreement with the experimental results and the SPP
predictions.  It appears therefore that in the far-zone, for both
silver/air and gold/air surfaces, far-field interferometry and
FDTD calculations show that the surviving long-range surface wave
is indeed the expected bound surface plasmon polariton.
Table\,\ref{table} summarizes the relevant parameters, far-field
interferometric measurements, and finite-difference-time-dependent
(FDTD) numerical simulations for gold and silver.
\begin{table}[htb]
\caption{\label{table}Summary of $\lambda_{\mathrm{surf}}$,
$n_{\mathrm{surf}}$ and $n_{\mathrm{spp}}$ determined from
far-field interferometric studies and FDTD simulations in gold and
silver}
\begin{ruledtabular}
\begin{tabular}{ccccc}
&$\epsilon_{\mathrm{m}}$&$\lambda_{\mathrm{surf}}$&$n_{\mathrm{surf}}$&$n_{\mathrm{spp}}$\\
Au,\,Ref.\,\cite{PresentExperimentalResults}&$-28.32$&$839\pm
6$~nm&$1.016\pm
0.004$&1.018\\
Au,\,Ref.\,\cite{PresentFDTDSimulations}&$-31.62$&$839$~nm&$1.016$&$1.016$\\
Ag,\,Ref.\,\cite{NearZoneResults1}&$-33.27$&$819\pm
8$~nm&$1.04\pm 0.01$&$1.015$\\
Ag,\,Ref.\,\cite{NearZoneResults2}&$-33.27$&$814\pm 8$~nm&$1.05\pm
0.01$&1.015\\
Ag,\,Ref.\,\cite{FarZoneSimulations}&$-33.98$&$837$~nm&$1.017$&$1.015$\\
\end{tabular}
\end{ruledtabular}
\end{table}
\subsection{Surface wave in the near-zone}
In the near-zone both experiment and numerical simulation show
that the surface wave deviates from pure SPP behavior.  The
effective propagation parameter $k_x$, originating near the slit
edge, appears greater than $k_{\mathrm{spp}}$ and evolves smoothly
to the bound mode over the near-zone interval of a few microns.
This behavior may be interpreted either in terms of initial
excitation of a composite evanescent surface ``wave packet" in
$k$-space at the slit edge\,\cite{LT04}, followed by subsequent
decay of all surface modes except the bound $k_{\mathrm{spp}}$
mode or in terms of detailed field matching at the boundaries
within the slit and near the slit edges\,\cite{XZM06}.  These two
points of view both invoke evanescent modes $k_x\geq
k_{\mathrm{spp}}$ in order to satisfy boundary conditions in the
vicinity of the slit edge, but standard wave-guide theory dictates
that only the SPP mode is stable against phonon coupling to the
bulk metal or to radiative decay.  We can estimate the the surface
distance over which the dissipation occurs by appealing to the
standard Drude model of a metal that expresses the frequency
dependence of the dielectric constant $\epsilon(\omega)$ in terms
of the bulk plasmon resonance $\omega_p$ and a damping constant
$\Gamma$.
\begin{equation}
\epsilon(\omega)=\epsilon_0\left(\epsilon_{\infty}-\frac{\omega_p^2}{\omega^2+i\Gamma\omega}\right)\label{Eq:
DrudeModel}
\end{equation}
In Eq.\,\ref{Eq: DrudeModel} $\epsilon_0$ is the permittivity of
free space and
$\epsilon_{\infty}=\epsilon(\omega\rightarrow\infty)$ is the
dimensionless infinite frequency limit of the dielectric constant.
The Drude model is based on a damped harmonic oscillator model of
an electron plasma in which the electrons oscillate about positive
ion centers with characteristic frequency $\omega_p$, subject to a
phenomenological damping rate $\Gamma$, normally assumed to be due
to electron-phonon coupling.  Values for $\Gamma$ are typically
$\sim 10^{14}$~s$^{-1}$, and in fact for Au the value is
$1.02\times 10^{14}$~s$^{-1}$.  For a wave propagating on the
surface with group velocity $\simeq 3\times 10^8$~ms$^{-1}$ the
expected decay length $\simeq 3~\mu$m, consistent with the
measurements and FDTD simulations.

In summary the picture that emerges from far-field interferometry
and FDTD simulation studies of these simple slit-groove structures
on silver and gold films is that in the near-zone of slit-groove
distances, on the order of a few wavelengths, the surface wave
consists of a composite of several evanescent modes all of which
dissipate within this near-zone.  Only the bound, stable SPP mode
survives into the far-zone, and in the studies reported here we
have observed essentially constant SPP amplitude out to 12~$\mu$m,
a distance limited only by our fabricated structures. Earlier
measurements\,\cite{GAV06a} indicate that absorption and surface
roughness scattering should permit propagation lengths as far as
$\sim 80-100~\mu$m.

\begin{acknowledgments}
Support from the Minist{\`e}re d{\'e}l{\'e}gu{\'e} {\`a} l'Enseignement sup{\'e}rieur et {\`a}
la Recherche under the programme
ACI-``Nanosciences-Nanotechnologies," the R{\'e}gion Midi-Pyr{\'e}n{\'e}es
[SFC/CR 02/22], and FASTNet [HPRN-CT-2002-00304]\,EU Research
Training Network, is gratefully acknowledged.  F.K. gratefullly
acknowledges support from the Deutsche Telekom Stiftung.
Facilities of the Caltech Kavli Nanoscience Institute are also
gratefully acknowledged.
\end{acknowledgments}

\end{document}